\documentclass[10pt,twocolumn,numbers]{article}

\usepackage{graphicx}
\usepackage{amssymb}
\usepackage{amsthm,amsmath,dsfont}
\usepackage{mathtools}

\usepackage{lineno}

\usepackage[colorlinks=true, citecolor=blue]{hyperref}
\usepackage[linesnumbered,lined,ruled,commentsnumbered]{algorithm2e}
\usepackage{algcompatible}
\usepackage{wrapfig}
\usepackage[numbers,sort&compress]{natbib}
\usepackage{comment}
\usepackage{longtable}
\usepackage{multirow}
\usepackage{bbding}
  
\usepackage[bottom]{footmisc}
\usepackage{booktabs}
\usepackage{pifont}% http://ctan.org/pkg/pifont
\usepackage{float}
\usepackage{subfigure}
\usepackage{wrapfig}
\usepackage{framed,xcolor}
\usepackage{newfloat}
\usepackage[xcolor,framemethod=TikZ]{mdframed}
\usepackage{mathpazo}
\usepackage{amsthm}
\usepackage{thmtools}
\usepackage{authblk}
\patchcmd{\thebibliography}{\section*{\refname}}{}{}{}
\usepackage{tikz}
\usetikzlibrary{arrows,shapes,chains,calc,patterns,decorations.pathreplacing}
\usepackage{stmaryrd}
\usepackage{fancyhdr}
\usepackage{subfigure}
\usepackage{comment}
\def\BibTeX{{\rm B\kern-.05em{\sc i\kern-.025em b}\kern-.08em
		T\kern-.1667em\lower.7ex\hbox{E}\kern-.125emX}}

\renewcommand{\headrulewidth}{2pt}

\newlength\FHoffset
\fancyheadoffset{\FHoffset}

\newlength\FHleft
\newlength\FHright

\newbox\FHline
\setbox\FHline=\hbox{\hsize=\paperwidth%
	\hspace*{\FHleft}%
	\rule{\dimexpr\headwidth-\FHleft-\FHright\relax}{\headrulewidth}\hspace*{\FHright}%
}

\newtheoremstyle{theoremdd}% name of the style to be used
{\topsep}% measure of space to leave above the theorem. E.g.: 3pt
{\topsep}% measure of space to leave below the theorem. E.g.: 3pt
{\itshape}% name of font to use in the body of the theorem
{0pt}% measure of space to indent
{\fontfamily{cmss}\selectfont\bfseries}% name of head font
{.}% punctuation between head and body
{ }% space after theorem head; " " = normal interword space
{\thmname{#1}\thmnumber{ #2}\thmnote{ (#3)}}

\theoremstyle{theoremdd}

\mdfdefinestyle{myStyle}{roundcorner=5pt,backgroundcolor=brown!20,linecolor=red!40!black,linewidth=1.5pt}

\usepackage{titlesec}
\titleformat*{\section}{\fontfamily{cmss}\selectfont\large\bfseries\color{red!40!black}}
\titleformat*{\subsection}{\fontfamily{cmss}\selectfont\normalsize\bfseries\color{red!40!black}}
\titleformat*{\subsubsection}{\fontfamily{cmss}\selectfont\normalsize\color{red!40!black}}
\usepackage[left=0.7 in, right=0.7 in, top=1.1 in, bottom=1.3 in]{geometry}
\graphicspath{{./Figures/}}

\renewcommand\abstractname{\fontfamily{cmss}\selectfont\normalsize\bfseries\color{red!40!black}\textbf{Abstract}}

\renewenvironment{abstract}{%
	\centering\small
	\list{}{\leftmargin1.5cm \rightmargin\leftmargin}
	\item\relax
	
	\begin{mdframed}[]
		\item[\hskip\labelsep\scshape\abstractname.]%
	}{%
	\end{mdframed}
	\endlist \par\bigskip
}
\makeatletter
\patchcmd{\@maketitle}{\LARGE \@title}{\fontfamily{cmss}\selectfont\LARGE\color{red!40!black}\@title}{}{}
\makeatother

\graphicspath{{./Figures/}}

\begin{document}

%	\begin{frontmatter}
		
		
		\title{Backpressure Control with Estimated Queue Lengths for Urban Network Traffic}
		
			

\author[1]{Li Li}
\author[2]{Victor Okoth}
\author[1,2]{Saif Eddin Jabari}

\affil[1]{New York University Tandon School of Engineering, Brooklyn NY, U.S.A.}
\affil[2]{New York University Abu Dhabi, Saadiyat Island, P.O. Box 129188, Abu Dhabi, U.A.E.}

\date{}

%\begin{mdframed}[style=myStyle]
%\end{mdframed}

\twocolumn[
\begin{@twocolumnfalse}
	
\maketitle	

\begin{abstract}
	Backpressure (BP) control was originally used for packet routing in communications networks. Since its first application to network traffic control, it has undergone different modifications to tailor it to traffic problems with promising results. Most of these BP variants are based on an assumption of perfect knowledge of traffic conditions throughout the network at all times, specifically the queue lengths (more accurately, the traffic volumes). However, it has been well established that accurate queue length information at signalized intersections is never available except in fully connected environments.  Although connected vehicle technologies are developing quickly, we are still far from a fully connected environment in the real world.   This paper test the effectiveness of BP control when incomplete or imperfect knowledge about traffic conditions is available. We combine BP control with a speed/density field estimation module suitable for a partially connected environment. We refer to the proposed system as a BP with estimated queue lengths (BP-EQ). We test the robustness of BP-EQ to varying levels of connected vehicle penetration, and we compared BP-EQ with the original BP (i.e., assuming accurate knowledge of traffic conditions), a real-world adaptive signal controller, and optimized fixed timing control using microscopic traffic simulation with field calibrated data. Our results show that with a connected vehicle penetration rate as little as 10\%, BP-EQ can outperform the adaptive controller and the fixed timing controller in terms of average delay, throughput, and maximum stopped queue lengths under high demand scenarios. 
	
	\medskip
	
	\textbf{\fontfamily{cmss}\selectfont\color{red!40!black} Keywords}: Backpressure control, traffic state interpolation.
\end{abstract}
\bigskip
\end{@twocolumnfalse}
]

%\begin{keyword}
%	Automated vehicles \sep cellular Automata \sep conditional random fields \sep stochastic traffic modeling \sep traffic state estimation \sep trajectory reconstruction
%\end{keyword}
		
%\end{frontmatter}
	
	
%	\linenumbers
	

\section{Introduction}
\label{S:intro}

Network traffic control can be classified into centralized and decentralized/distributed control \citep{manolis2018centralised}. While centralized control can coordinate adjacent intersections and achieve system optimality (in theory), in practice they are not scalable computationally and have limited applicability.  In contrast, decentralized controller distribute computations to the network intersections, and are hence scalable to large networks.

Backpressure (BP) based techniques, as one of the distributed traffic controllers, have attracted a lot of attention from the researchers in recent years because of the distinct advantages they offer: 1) they are provably network stabilizing and 2) they require no knowledge about the traffic arrival rates. BP was first brought in traffic network control from the communications networks theory independently by Wongpiromsarn et al. \citep{wongpiromsarn2012distributed} and Varaiya \citep{varaiya2013max}. In BP, each intersection collects its local queue length information, and makes decisions on which phase to activate every time step. Specifically, queue lengths are used to calculate the pressure of each phase and the phase with the maximum pressure is then chosen as the phase to be activated. This is why BP control is also called max-pressure control sometimes, e.g., \citep{varaiya2013max, kouvelas2014maximum, hosseini2016comparison, sun2018simulation, mercader2019experimental,mercader2020max}.

The accuracy of the queue lengths used in BP control can impact its performance. To avoid confusion, by ``\textit{queue}'', what is meant in this paper is the total number of vehicles in a road segment (i.e., along a network link).  This is consistent with the definition used by queuing theorists, the vast majority of BP models in traffic control and, more importantly, it is consistent with the way queues are treated in the dynamical models used to produce the network-wide performance guarantees.  It has been established that this is a deficiency in BP models since control decisions should distinguish stopped or ``queued'' vehicles (near the stoplines) from those moving freely near the entrances of the network links \citep{li2019position}. Many variants of BP have appeared in the literature (e.g. \citep{kouvelas2014maximum, gregoire2014back, gregoire2014capacity, le2015decentralized, sun2018simulation, liu2018back, li2019position, mercader2019experimental,mercader2020max}), each tuning BP to different traffic nuances, most of these approaches  assume that the controller has exact knowledge of the queue lengths (e.g. by assuming a fully connected environment), and focused on how to use the known queue lengths to calculate the pressure in different ways. Mercader et al. \citep{mercader2019experimental,mercader2020max} pointed out that queue lengths at signalized intersections are hard to estimate in today's traffic environment and even in a partial connected vehicles (CVs) environment expected in the near future. However, they have not attempted to investigate how BP performs with missing and/or inaccurate queue length information and proposed the use of travel times (which are easier to estimate) instead of queue lengths to calculate the pressure \citep{mercader2019experimental,mercader2020max}.

To the best of the authors' knowledge, the robustness of BP to inaccuracies in queue lengths has not been investigated.  Most BP based papers simply assume a fully connected environment.  While recent years have seen great developments towards such an environment, we are still far from realizing it in practice.  The performance of BP in the absence of detailed/accurate queue length information is the subject of this paper.

We assume a partially connected setting, in which some of the vehicles in the network can communicate with the infrastructure, specifically intersection controllers.  We propose a method to feed the data provided by the connected vehicles in the network to the BP controller. In brief, the basic idea includes three steps: 1) use a speed interpolation technique to create a real time speed map estimate of the road using the data that is available; 2) convert the speeds to densities using a speed-density relationship; and 3) calculate the queue length from density and use it for pressure computation in BP. The methodology is tested using microscopic traffic simulation of a calibrated isolated intersection and a traffic network from the real world. Comparisons are made with a real-world implementation of an adaptive controller, an optimized fixed timing controller, and BP with perfect knowledge.

The rest of the paper is organized as follows: the following section gives a detailed literature review on BP based control for traffic networks, the third section introduces the proposed BP with estimated queue length (BP-EQ) approach, followed by simulation experiments and results in the fourth section. The last section concludes the paper.

\section{Background and literature review}
\label{S:literature}

BP was initially proposed by Tassiulas and Ephremides \citep{tassiulas1992stability} as a solution for the packet routing problem in communications networks. It can be applied in a completely distributed manner while ensuring network-wide stability of the queues for all possible mean arrival rates that lie within the capacity region of the network.  Another significant advantage of BP is that it requires no knowledge of network demands.  It was first introduced into network traffic signal control by Wongpiromsarn et al. \citep{wongpiromsarn2012distributed} and Varaiya \citep{varaiya2013max} independently about 7 years ago. Since then, it has seen many extensions geared towards adapting BP to various features of road network traffic.

The original traffic signal network BP in \citep{wongpiromsarn2012distributed} and \citep{varaiya2013max} calculates the difference between queue lengths of upstream and downstream links of an intersection movement and take the product of this queue difference and the saturation flow rate of the movement as the pressure of that movement. The pressure of a traffic signal phase (a combination of non-conflicting movements) is then defined as the sum of the pressures of all movements that belong to the phase.  At the end of each time slot, each intersection collects its local queue length information, and chooses the phase with the maximum pressure to activate in the next time slot. The original BP has the following properties, which are not suitable for traffic control:

\begin{enumerate}
	\item It involves frequent updates (every time step).
	\item It assumes infinite spatial capacity of the road to prove network stability.
	\item It assumes predetermined routing decisions for vehicles, which will not dynamically change according to real-time traffic conditions and signal states.
	\item It used queue lengths for pressure calculations.
	\item It requires the knowledge of saturation flow rates (and turning ratios in \citep{varaiya2013max}) for each movement.
\end{enumerate}

While the first property is suitable for packets to be routed and re-routed at a fine cadence (e.g. every 1 s), this can translate to aggressive signal switching, which is not suitable for traffic control. To overcome this issue, Varaiya et al. \citep{kouvelas2014maximum} introduced a cyclic BP which has a fixed phase sequence and predetermined cycle length. At the beginning of each cycle, the intersection assigns the effective green of each phase in proportional to their corresponding pressure. However, Varaiya et al. did not provide a stability proof for this cyclic BP, but showed that it has comparable performance with original BP in simulation experiments. Following the work of \citep{kouvelas2014maximum}, Le et al. \citep{le2015decentralized} proposed a similar cyclic phase BP and provided a proof of stability. Their proof demonstrated that BP does not require accurate turning ratios to guarantee stability, and that an unbiased estimator of the turning ratio is sufficient. Their simulation experiments indicate that the cyclic BP outperforms the original BP.  In light of the work by \citep{kouvelas2014maximum} and \citep{le2015decentralized}, Taale et al. \citep{taale2015integrated} and Sun et al. \citep{sun2018simulation} tested both a time-slotted BP and a cyclic BP in their experiments, and interestingly, they found that the time-slotted BP works better than the cyclic BP in general.  More recently, \citep{rey2019blue} proposed a BP that can accommodate both human-driven and autonomous vehicles.

The main issue with the second property is loss of work conservation \citep{gregoire2014capacity} and loss of theoretically provable stability of BP in a real traffic networks where all physical roads have limited capacities \citep{xiao2014pressure, zaidi2015traffic, zaidi2016back}. Work conservation was addressed in a heuristic way in \citep{gregoire2014capacity} but no guarantees of stability were delivered with their heuristic.  An alternative pressure formulation was proposed in \citep{xiao2014pressure} to address work conservation under finite spatial capacity with guarantees of stability. However, their approach requires knowledge of non-local queue lengths in addition to the local queue lengths, which makes it impractical. Recently, \citep{li2019position} proposed a position weighted backpressure that incorporates the spatial distribution of vehicles in the pressure calculation. They theoretically proved the stability of their approach under the finite network capacities for the first time, without requiring any non-local traffic information.

Recent literature has addressed the third property by combining traffic control with vehicle routing using BP techniques. Among them, Chai et al. \citep{chai2017dynamic} tested BP with dynamic vehicle routing, and found that enabling vehicle re-routing in the network can reduce vehicle travel times as well as queue lengths at intersections. Their approach applies BP only for signal control. Taale et al. \citep{taale2015integrated} used BP for both signal control and vehicle routing. They defined the pressure for each route and vehicles make dynamic routing decisions based on the calculated pressure. However, in both \citep{chai2017dynamic} and \citep{taale2015integrated}, signal optimization and vehicle routing decisions are made separately. Other papers sought to optimize signal control and vehicle routing simultaneously \citep{zaidi2015traffic, gregoire2016back, zaidi2016back, le2017utility, liu2018back}. They typically employ multi-commodity queues, where each destination in the network serves as a commodity. The intersection controller then uses a multi-commodity BP to determine the vehicle routing along with the signal phasing. Such a BP with adaptive routing would require not only the vehicles and intersections to be connected to communicate the destinations and routing decisions, but also require all vehicles to follow the routing decisions made by the intersection controller.  This was improved in \citep{gregoire2016back}, where network stabilizing BP with adaptive routing was proposed which only provides route guidance to a subset of vehicles in the system. Le et al. \citep{le2017utility} considered vehicle compliance with the routing in their approach.

Using queue length for pressure calculation (property 4) would lead to a fairness issue caused by the well-known ``last packet problem'' \citep{wu2017delay}, where a vehicle in a short queue that does not grow for a long time will suffer a long wait time as a queue-based BP will always favor movements with long queues.  Such issues also exist in communications network \citep{ji2012delay}, and one of the solutions is to use delay of the head-of-line (HOL) packet rather than queue length to compute pressure, which results in a delay-based BP. Wu et al. \citep{wu2017delay} adopted the idea from \citep{ji2012delay} and proposed a HOL-delay based BP and a queue-delay-mixed BP for a single intersection. They proved the stability of both BP policies, and demonstrated from simulation that delay-based BP has the same stability region as a queue-based BP, while ensuring fairness. However, they failed to find a way to decompose the delay-based BP (it is not decentralized), which makes their controller scale poorly to large networks. Based on the work of \citep{wu2017delay}, Yen et al. \citep{yen2018falsified} tested the robustness of queue-based BP and HOL-delay based BP to the falsified data attacks. The attackers are simulated as vehicles that spoof their arrival times. The simulation results show that the HOL-delay based BP is more vulnerable to time spoofing attacks.

Another concern with the fourth property is the requirement that accurate queue length information is available. Considering the difficulties in queue length estimation at signalized intersections in today's road networks, Mercader et al. \citep{mercader2019experimental,mercader2020max} put forward a travel time based BP. Instead of using queue lengths as the input for calculating pressure, they used the travel times because 1) travel times are easier to obtain or estimate than queue lengths, and 2) travel times tend to diverge as queue sizes approach the spatial capacity of roads, hence it is inherently capacity aware.  The authors implemented this travel time based BP policy in a real intersection in Jerusalem by installing four Bluetooth detectors in field to collect the data from detectable Bluetooth devices such as smartphones and in-vehicle audio systems. They found that their controller is superior to the optimized fixed control scheme in field application. This is the first (and so far the only) implementation of BP in the real world.

Regarding the fifth property, it is unrealistic to assume perfect knowledge of turning ratios in today's road networks.  Gregoire et al. \citep{gregoire2014back} put forward a modified BP controller with unknown routing rates. They proved that the modified BP can guarantee stability in heavily congested conditions, but not all traffic conditions.  Xiao et al. \citep{xiao2015throughput, xiao2015further} proposed an extended BP using online estimators with fading memories to estimate queue lengths, turning ratios, and saturation flow rates in real time using measured values from detectors. They then constructed a new pressure formulation using the values from the estimators. Given that the estimations are unbiased, they proved the stability of the extended BP under some assumptions. However, they did not specify how to measure the queues/turning ratios/saturation rates using detectors.

There are other notable BP based control algorithms developed for traffic control.  For example, Levin and Boyles \citep{levin2017pressure} used BP techniques in reservation-based intersection control for connected automated vehicles; Hao et al. \citep{hao2019distributed} proposed a cooperative BP, which uses a new pressure calculation method that considers signal control of downstream intersections to achieve coordination between adjacent intersections.

On the one hand, queue lengths are basic inputs for BP and are assumed to be known with accuracy in most of the existing studies \citep{wongpiromsarn2012distributed, varaiya2013max, kouvelas2014maximum, gregoire2014back, xiao2014pressure, le2015decentralized, zaidi2015traffic, gregoire2014capacity, taale2015integrated, gregoire2016back, zaidi2016back, hosseini2016comparison, chai2017dynamic, levin2017pressure, le2017utility, yen2018falsified, sun2018simulation, liu2018back, hao2019distributed, li2019position}.  On the other hand, the only BP controller that was tested in the real world \citep{mercader2019experimental,mercader2020max} used travel times instead of queue lengths for pressure calculation due to the difficulty to obtain exact queue length. The only papers that considered the estimation of queue lengths in BP control \citep{xiao2015throughput, xiao2015further} propose an estimator which takes measured queue as inputs without specifying how to measure queues using detectors or probe vehicles (typical real-world instrumentation). There are no study about the robustness of BP to the accuracy of the estimated queue lengths.  The main purpose of this paper is to fill this research gap.

\section{BP-EQ Methodology}
\label{S:model}

\subsection{Space Discretization}
\begin{figure*}[ht!]%
	\centering
	\resizebox{0.75\textwidth}{!}{%
		\includegraphics{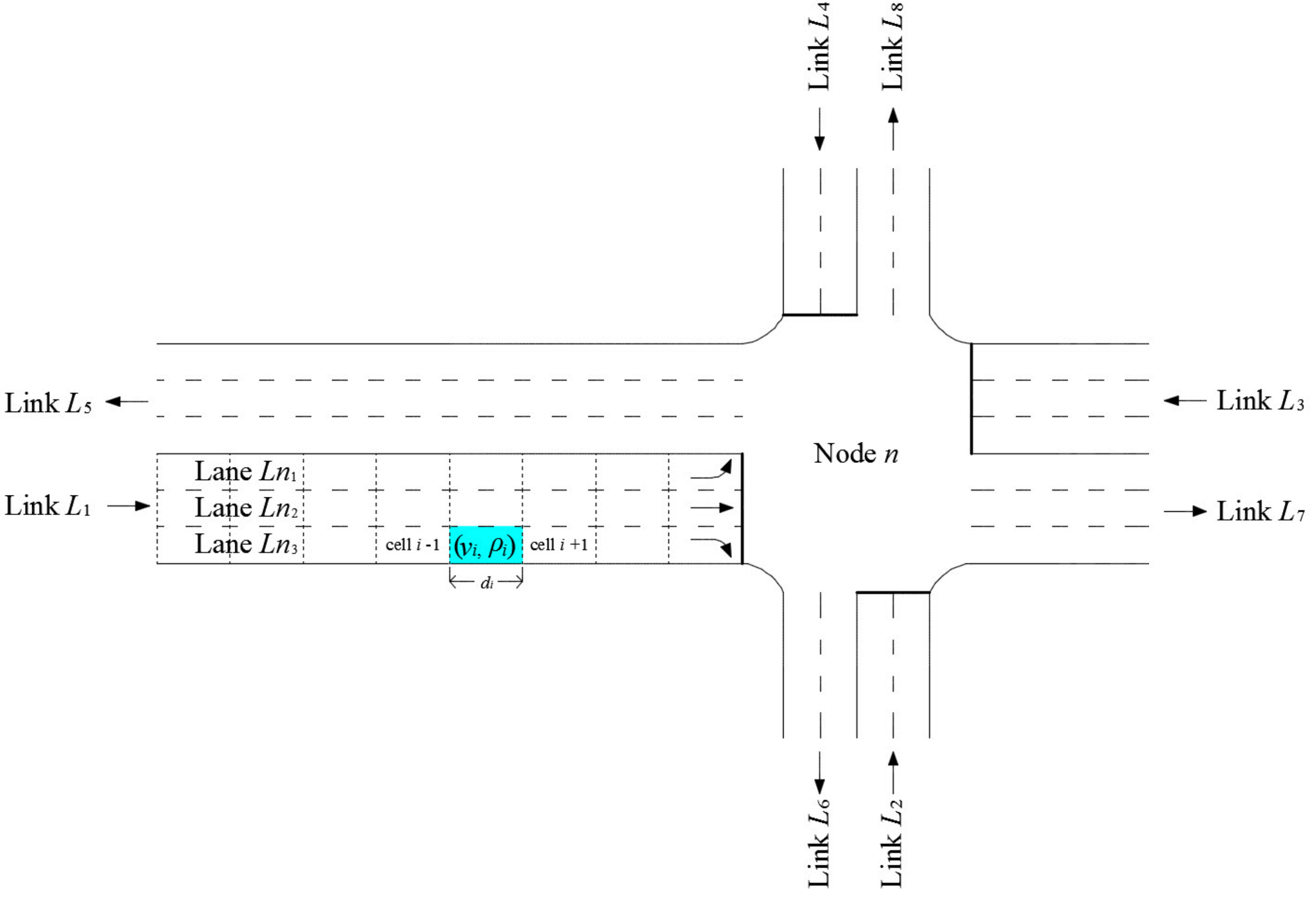}}	
	\caption{Road discretization of an intersection.}
	\label{F:road_discrete}%
\end{figure*}

As shown in Fig.~\ref{F:road_discrete}, we discretize the road into small cells of equal length and assume that the traffic density within each cell is constant. BP-EQ for a partially connected environment involves three steps:
\begin{enumerate}
	\item First, estimate the speed in each cell using data from connected vehicle (CV) probes.
	\item Then, calculate cell traffic density from estimated cell speed using a speed-density relation.
	\item Finally, calculate the queue sizes from the estimated cell densities and use them as input into BP.
\end{enumerate}
These steps are described in more detail next.

\subsection{Estimate Speed}
Treiber et al. \citep{treiber2013traffic} proposed a traffic state reconstruction method using interpolation techniques. This can also be used for real time
\begin{figure*}[ht!]
	\centering
	\subfigure[Penetration rate = 10\%]{\resizebox{0.4\textwidth}{!}{
			\includegraphics{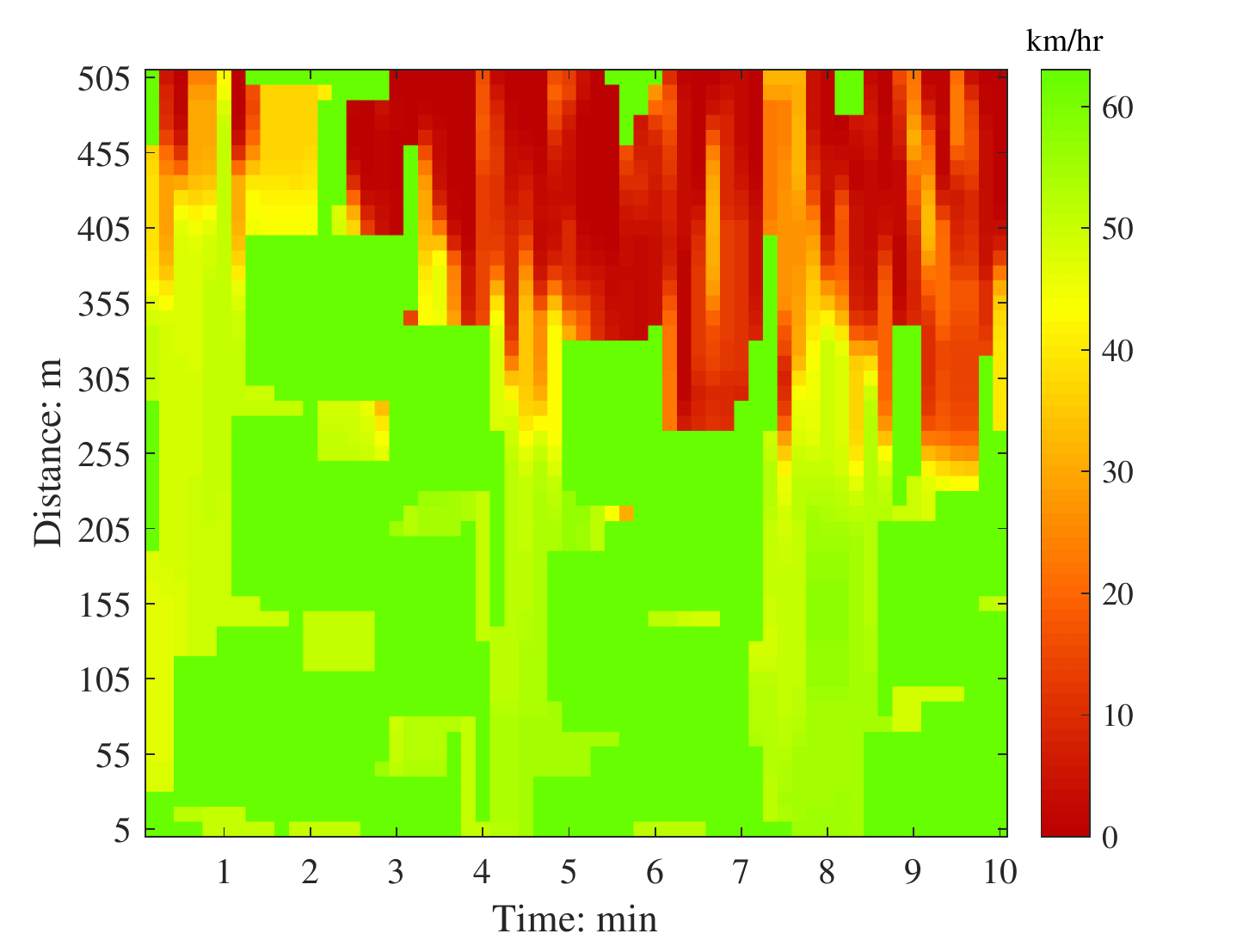}}
		\label{F:HM10}}
	\subfigure[Penetration rate = 20\%]{\resizebox{0.4\textwidth}{!}{
			\includegraphics{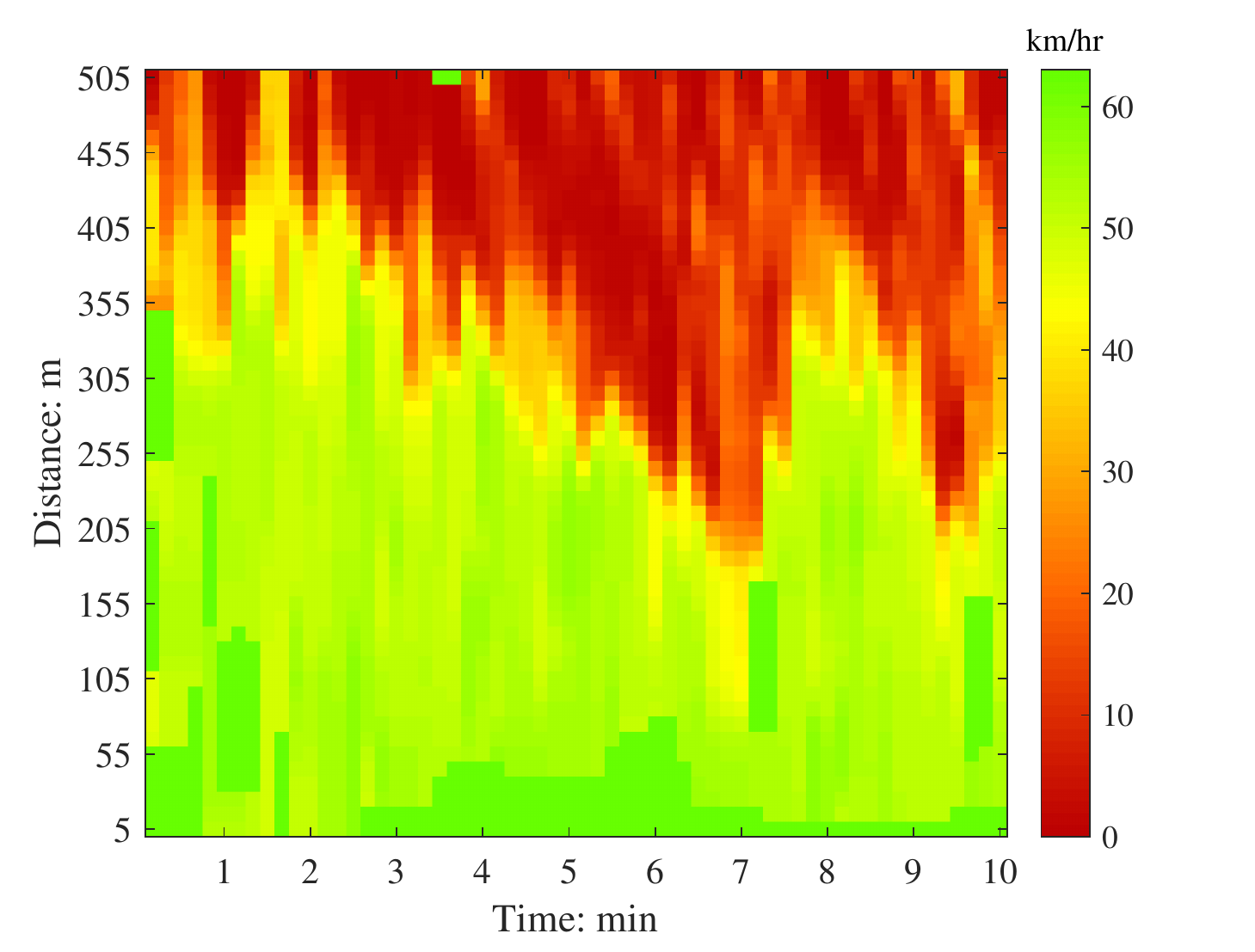}}
		\label{F:HM30}} \\
	\subfigure[Penetration rate = 30\%]{\resizebox{0.4\textwidth}{!}{
			\includegraphics{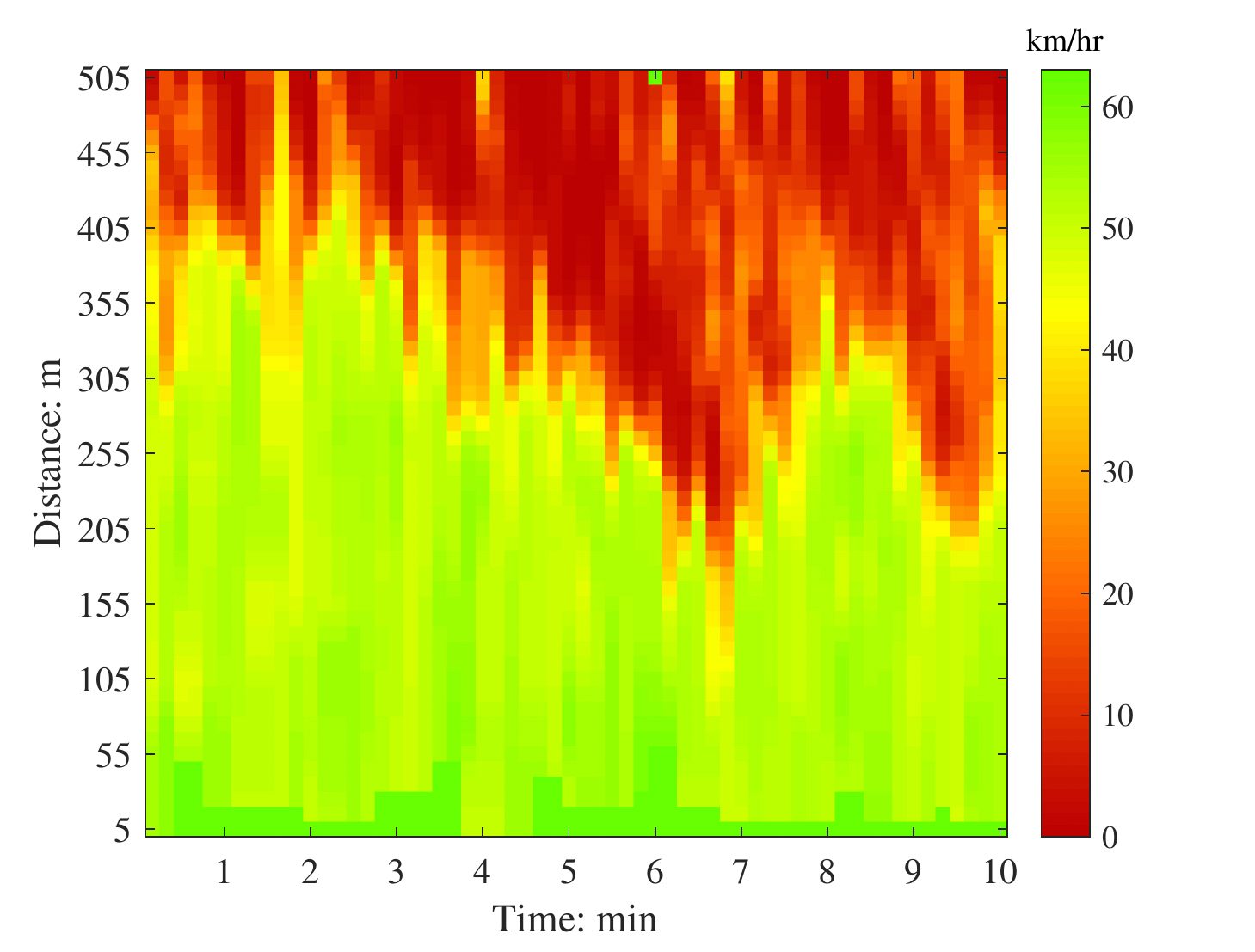}}
		\label{F:HM50}}
	\subfigure[Penetration rate = 100\%]{\resizebox{0.4\textwidth}{!}{
			\includegraphics{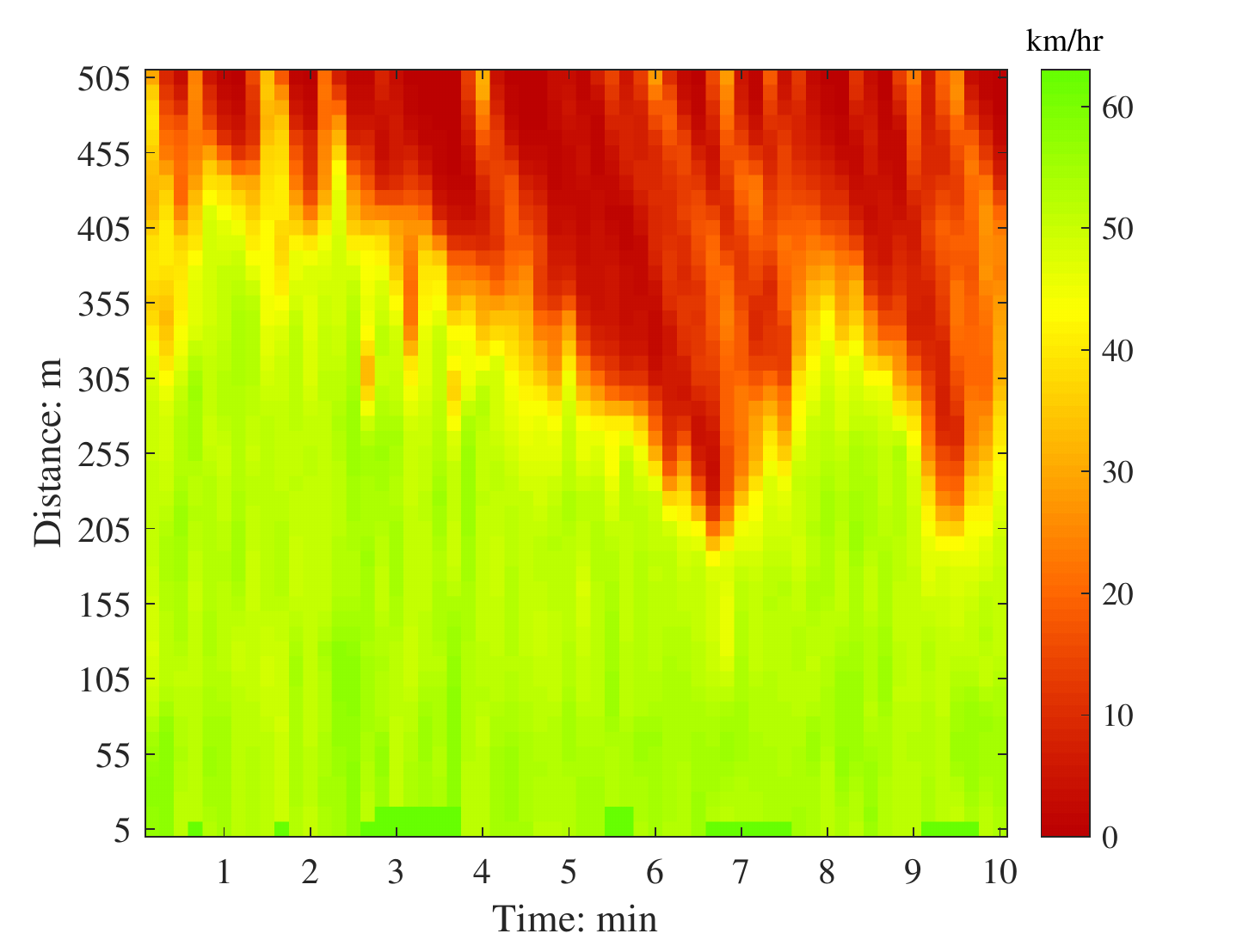}}
		\label{F:HM100}} \\
	\caption{Estimated spatial-temporal speed profiles at different penetration levels of connected vehicles.}
	\label{F:speed}
\end{figure*}
traffic estimation. The basic idea is to calculate the speed at each location by weighing available (both current and historical) speed data according to their collection locations and time stamps. Specifically, when estimating the speed at position-time stamp pair $(x_i,t_i)$, to determine the weight of a speed measurement $v_k$ collected at position-time stamp pair $(x_k, t_k)$, a kernel $\phi_0$ is used, as shown in \eqref{E:phi_0}. 
\begin{equation}
	\phi_0(x_i-x_k, ~t_i-t_k) = \exp \Big(-\frac{|x_i-x_k|}{\sigma} - \frac{|t_i-t_k|}{\tau} \Big), \label{E:phi_0}
\end{equation}
where position is the distance from the upstream end of the link. 
As recommended in \citep{treiber2013traffic}, a good value for $\tau$ is half of the data collection interval, and a good value for $\sigma$ is half of the average distance between data collection points. The weight of speed data $v_k$ when estimating the speed at $(x_i, t_i)$ is then calculated as:
\begin{equation}
	w_k(x_i,t_i) = \frac{\phi_0(x_i-x_k,t_i-t_k)}{Z(x_i,t_i)}, \label{E:weight}
\end{equation}
where the normalization denominator $Z(x_i, t_i)$ is the sum of all kernels for $(x_i, t_i)$:
\begin{equation}
	Z(x_i,t_i) = \sum_k \phi_0(x_i-x_k,t_i-t_k). \label{E:denominator}
\end{equation}
The speed at $(x_i, t_i)$ is then estimated as:
\begin{equation}
	v(x_i,t_i) = \sum_k w_k(x_i,t_i)v_k. \label{E:speed}
\end{equation}
For a fixed $(x_i, t_i)$, $w_k(x_i,t_i)$ decreases as $|x_i-x_k|$ increase and also decreases as $|t_i-t_k|$ increases. The farther data point $(x_k,t_k)$ is from the estimation location (in both space and time), the lower its weight will be in the estimated speed. We only use data within a lane to estimate the speed of points in that lane for simplicity in this paper. For example, when we are estimating the speed $v_i$ of cell $i$ in Fig.~\ref{F:road_discrete}, we only utilize the speed data from Lane $Ln_3$ of Link $L_1$. This simplification is reasonable for speed estimation at signalized intersections since different lanes may serve different movements (e.g. left-turn lane and through lane) and speeds in different lanes can vary dramatically as a result of signal control, so that using data from other lanes can introduce errors in the estimated speeds.

In a partially connected environments, the data collection points are floating CVs. Those CVs send their speeds and locations to the intersections, the intersections then store the historical data (within a certain time horizon $T$), and use the data to estimate the speed of each cell in Fig.~\ref{F:road_discrete}. We collect data every 10 s, hence $\tau$ is set to 5 s. $\sigma$ is calibrated to 20 m. Fig.~\ref{F:HM10} to Fig.~\ref{F:HM50} show the estimation results for CV penetration rates of 10\%, 20\%, and 30\%, respectively; Fig.~\ref{F:HM100} is for a penetration rate of 100\% (fully connected environment), which is regarded as the ground truth.

In this experiment, the road segment is 500 m long, with an intersection at the end of the road. The free flow speed is 60 km/h. We descritize the road into cells of length 10 m. For each cell on the road, we assume the density is constant. We take the central point as the position $x_i$ of cell $i$ and then calculate the density at that point. Since the traffic state of this road is changing fast due to the downstream intersection, we did not use data points from times that are too far in the past and set the time horizon $T$ to 4 time steps (which is 40 s in our case), meaning that we only use the speed data that are collected within the past 40 s. The vertical axis of Fig.~\ref{F:speed} is the distance from the starting point of the road. As we can see in Fig.~\ref{F:speed}, there is obvious formation and dissipation of queues on the road, caused by the traffic light at the downstream intersection. The traffic state near the stop line remained congested because the downstream intersection continued to be congested (i.e., due to spillback). The queue size was reduced during the green time, but only some of the vehicles in queue were served during the green. When the penetration rate is 30\%, the estimated speed map is very close to the ground truth map. This suggests that the interpolation techniques we are using perform quite well. Note that the sudden changes in speed (from red to green) near the downstream intersection are caused by the absence of data (from CVs) at the stop line. The estimation technique used \citep{treiber2013traffic} interpolates speeds between pairs of data points, for positions that are not sandwiched in between two data points, we assumed free flowing traffic conditions. Hence when the CV penetration rate is low, we have a higher probability of having no CVs right before the stop line. This is why we see more sudden changes in Fig.~\ref{F:speed}(a) than in Fig.~\ref{F:speed}(b) and Fig.~\ref{F:speed}(c).

It is worth mentioning that there also exist many other sophisticated queue length/size estimation and traffic state estimation methods in the literature \citep{liu2009real,wu2010identification,wu2011empirical,wu2011shockwave,jabari2019learning,benkraouda2020trafficState,benkraouda2020trafficData}. We chose \citep{treiber2013traffic} in this paper because of its simplicity and its ability to estimate the traffic state along the entire length of the road (not just the segment with queued vehicles near the stop line) under both light and heavy traffic conditions.

\subsection{Speed to Density} \label{ss:v2rho}
\begin{figure}[h!]
	\centering
	\subfigure[Speed density relation]
	{%
		\centering
		\includegraphics[width=.45\textwidth]{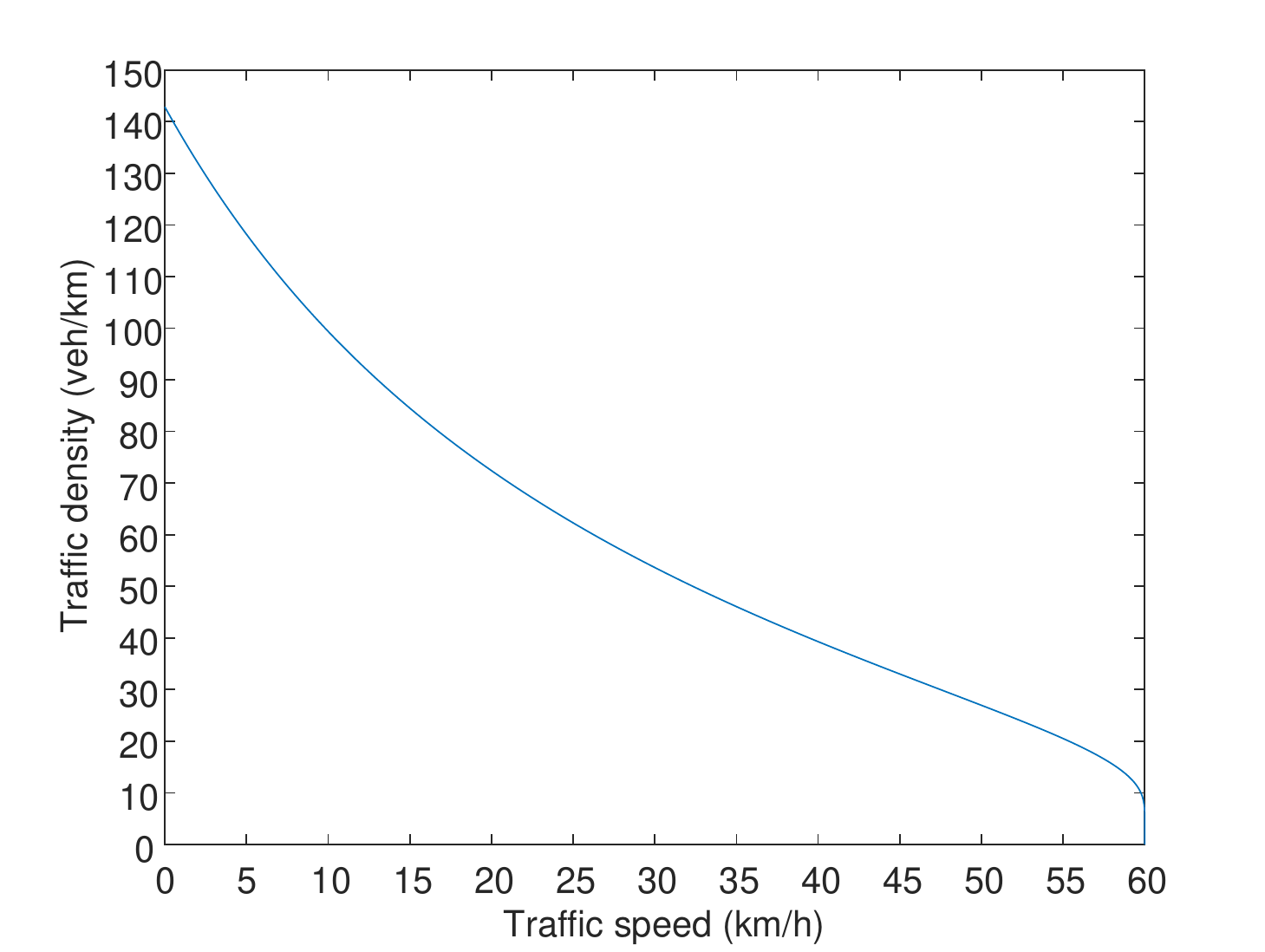}
		\label{F:distance}
	}%

	\subfigure[Density heat map]{%
		\centering
		\includegraphics[width=.45\textwidth]{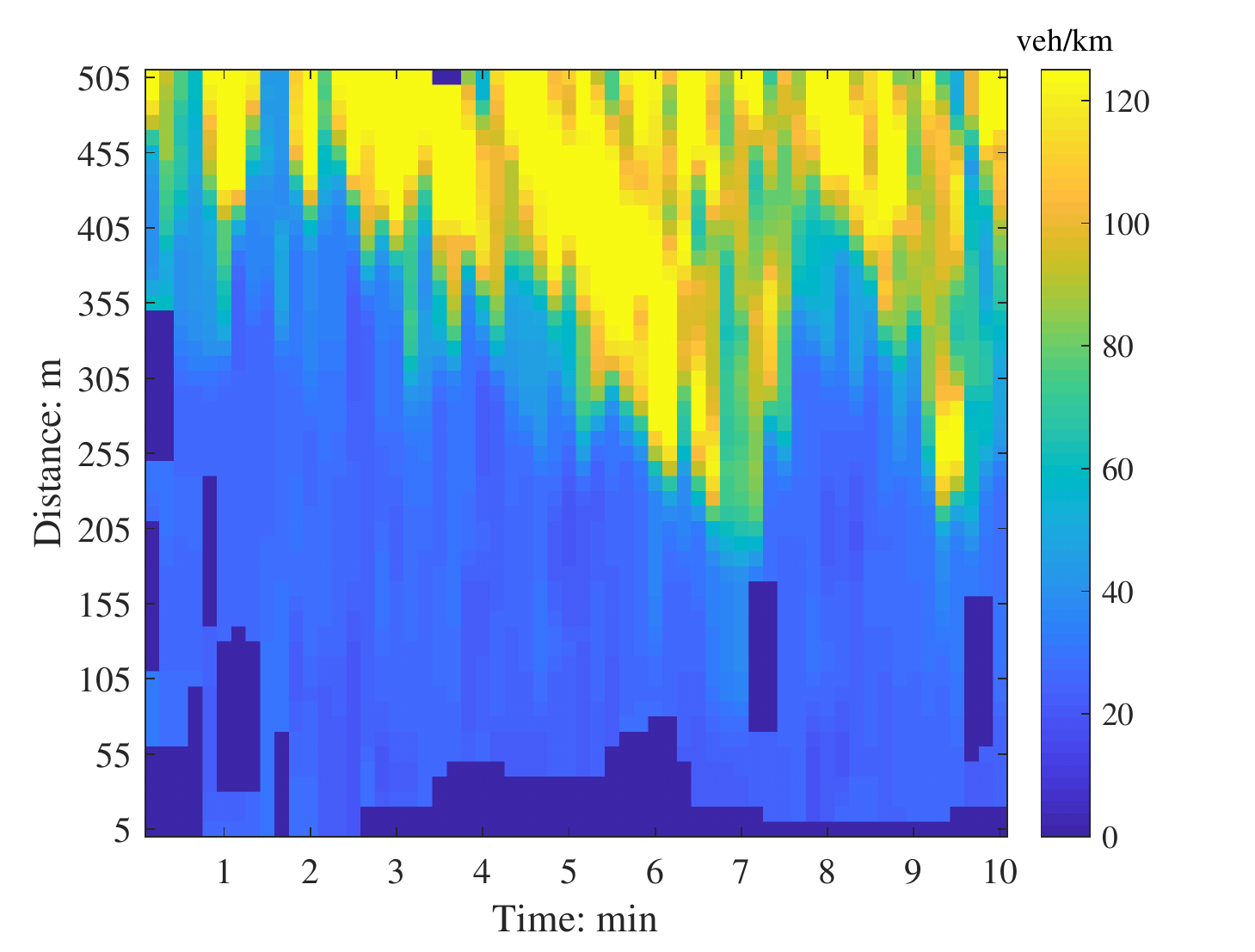}
		\label{F:time}
	}%
	\caption{Density-speed relationship.}
	\label{F:speed_density}
\end{figure}

We use the Newell-Franklin speed-density relation \citep{newell1961nonlinear, franklin1961structure} to estimate traffic densities from the estimated speeds in each cell. The Newell-Franklin relation has all the desirable feature that it is invertible, in other words, each speed is associated with one density. In addition to modeling convenience (invertibility), It also offers theoretical and empirically-proven advantages.  We refer to the detailed analyses in \citep{del1995functionali,del1995functionalii} for more details.  The relation is given by:
\begin{equation}
	\rho(x_i,t_i) = \frac{\rho_{\mathrm{jam}}}{1-\frac{v_{\mathrm{f}}}{w}\text{ln}(1-\frac{v(x_i,t_i)}{v_{\mathrm{f}}})}, \label{E:rho_v}
\end{equation}
where $\rho_{\mathrm{jam}}$ is the jam density, $v_{\mathrm{f}}$ is the free flow speed, $w$ is the shock wave speed. By setting $v_{\mathrm{f}}$ = 60 km/h, $w$ = 25 km/h, and $\rho_{\mathrm{jam}}$ = 143 veh/km, we have the density-speed relation shown in Fig.~\ref{F:distance}.
When the speed is 0, the density is equal to jam density, and the density decreases as the speed increases. The density at free flow speed is 0. Fig.~\ref{F:speed_density}(b) shows the density map corresponding to the speed map in Fig.~\ref{F:speed} (b) using the density-speed relation in \eqref{E:rho_v}.

It is worth mentioning that we chose the Newell-Franklin speed-density relationship because it is invertible.  A problem that one encounters with \emph{any} equilibrium relation is that higher speeds are associated with a range of traffic densities.  This means that estimates of traffic density in free-flow conditions are always prone to error.  However, we still estimate low traffic densities (free-flow conditions).  Moreover, the range of low traffic densities that are associated with free-flow speeds is narrow when compared to the range of traffic densities that are super-critical.  We utilize this conversion in our paper as a simplification and for purposes of testing the performance of BP in situations where there is incomplete information.  The inversion can be improved by utilizing probabilistic models \citep{jabari2014probabilistic,jabari2016sensor}, statistical/learning methods for parameter estimation \citep{jabari2012,jabari2013stochastic,seo2017traffic,jabari2018stochastic,zheng2018traffic,jabari2019learning}, and online update techniques \citep{dilip2017sparse,dilip2018systems,jabari2019sparse}.

\subsection{BP-EQ Control}

We use a simple BP control algorithm \citep{wongpiromsarn2012distributed} in this paper so as to focus on the impact of queue length estimation accuracy on BP performance, and leave the impact of turning ratio and separate queue estimation accuracy to future research.

\begin{figure}[ht!]
	\centering
	\resizebox{0.4\textwidth}{!}{%
		\includegraphics{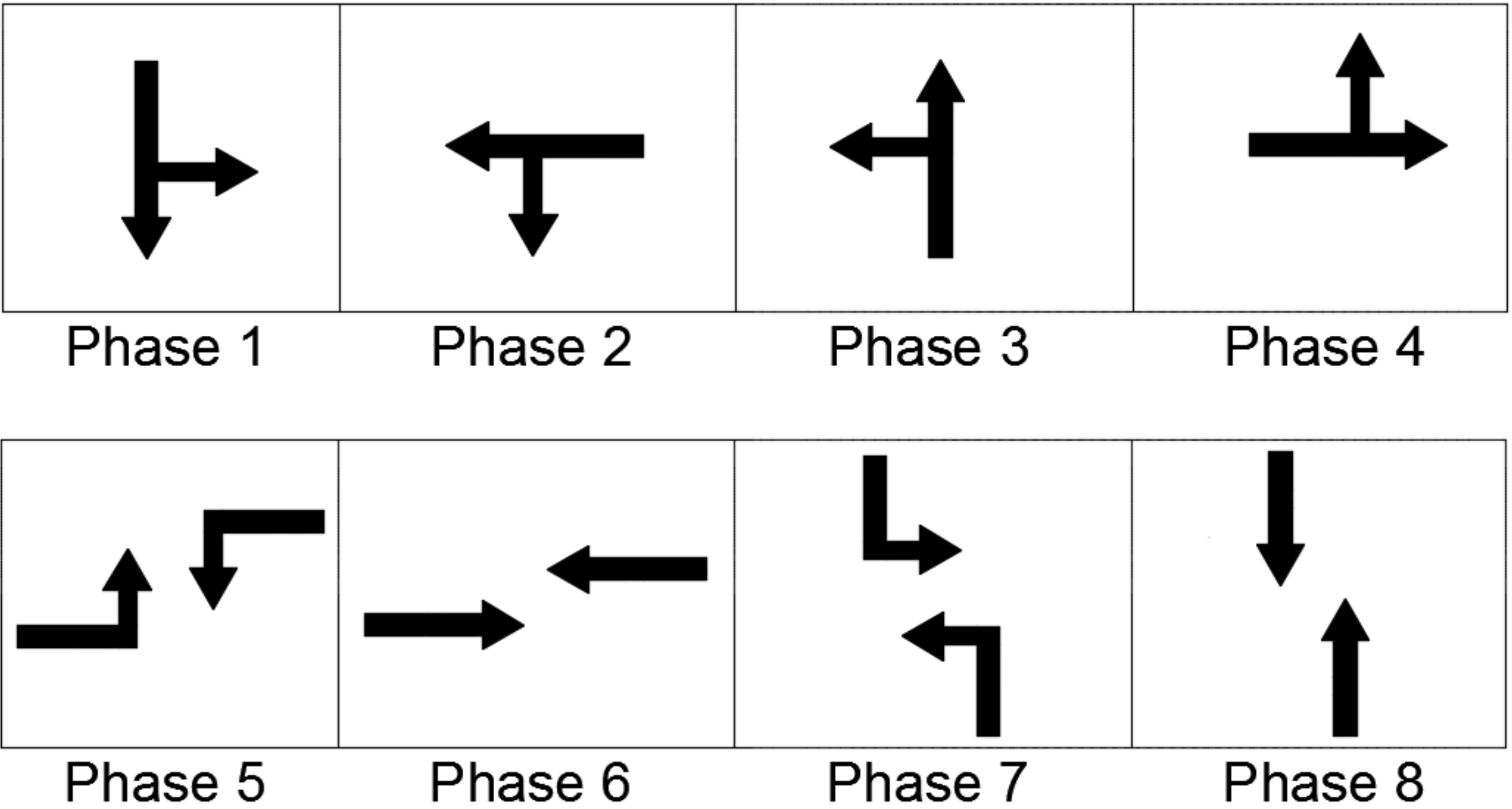}}
	\caption{\textbf{Phase set for a four-leg intersection}} \label{F:phases}
\end{figure}

When applying BP, one needs to first list all possible phases for the intersection. A phase here refers to a combination of non-conflicted movements. Fig.~\ref{F:phases} shows an example of all possible phases for a four-leg intersection such as the one shown in Fig.~\ref{F:road_discrete}. Note that we do not consider the right-turn movements in Fig.~\ref{F:phases} since right-turn traffic is usually not controlled by a traffic light. We also do not consider a single movement as a phase since we want to maximize the efficiency of each phase.

Let $\mathcal{P}_n$ denotes the set of all possible phases at intersection $n$. The BP algorithm in \citep{wongpiromsarn2012distributed} says that at the beginning of each time slot $t$, intersection $n$ should choose the phase $p_n(t)$ that maximizes the pressure:
\begin{equation}
	p_n(t) \equiv \text{arg} \max_{p_n \in \mathcal{P}_n} \sum_{(L_a,L_b)} W_{(L_a,L_b)}(t)\mu_{(L_a,L_b)}(p_n), \label{E:pn}
\end{equation}
where $(L_a, L_b)$ denotes a movement from Link $L_a$ to Link $L_b$, $\mu_{(L_a, L_b)}(p_n)$ is the saturation flow rate of movement $(L_a, L_b)$ if phase $p_n$ is activated, $W_{(L_a, L_b)}(t)$ is the weight associated with movement $(L_a, L_b)$ in time slot $t$. The weight is calculated as
\begin{equation}
	W_{(L_a,L_b)}(t) \equiv Q_{L_a}(t) - Q_{L_b}(t), \label{E:weightW}
\end{equation}
where $Q_{L_a}(t)$ is the queue length (number of vehicles) on Link $L_a$ at the end of time slot $t-1$. Once the cell densities have been calculated using \eqref{E:rho_v}, $Q_{L_a} (t)$ is calculated as
\begin{equation}
	Q_{L_a}(t) = \sum_{i \in L_a} \rho_i d_i, \label{E:Q}
\end{equation}
where $\rho_i$ is the traffic density of cell $i$ and $d_i$ is the length of cell $i$.
Substituting \eqref{E:weightW} and \eqref{E:Q} into \eqref{E:pn}, we have
\begin{equation}
	p_n(t) \equiv \text{arg} \max_{p_n \in \mathcal{P}_n} \sum_{(L_a,L_b)} \Big(\sum_{i \in L_a} \rho_i d_i - \sum_{i \in L_b} \rho_i d_i\Big) \mu_{(L_a,L_b)}(p_n). \label{E:pn_integrate}
\end{equation}
In \eqref{E:pn_integrate}, $\mu_{(L_a, L_b)}(p_n)$ can be calculated as
\begin{equation}
	\mu_{(L_a,L_b)}(p_n) = \begin{cases}
		0 & (L_a, L_b) \notin p_n \\
		\mu \cdot x_{(L_a,L_b)} \cdot f_t \cdot \Delta t / 3600 & (L_a, L_b) \in p_n \\
	\end{cases},
	\label{E:mu}
\end{equation}
where $\mu$ is the saturation flow rate of one through lane (a default value frequently used is 1800 veh/h), $x_{(L_a, L_b)}$ is the number of lanes used by movement $(L_a, L_b)$, $f_t$ is the turning adjustment factor of movement $(L_a, L_b)$ (a recommended value for left-turn movement is 0.714 \citep{roess2011traffic}), and $\Delta t$ is the duration of each time slot in units of seconds.

\section{Experiments}
\label{S:experiments}

\subsection{Isolated Intersection}
\begin{figure}[ht!]%
	\centering
	\resizebox{0.45\textwidth}{!}{%
		\includegraphics{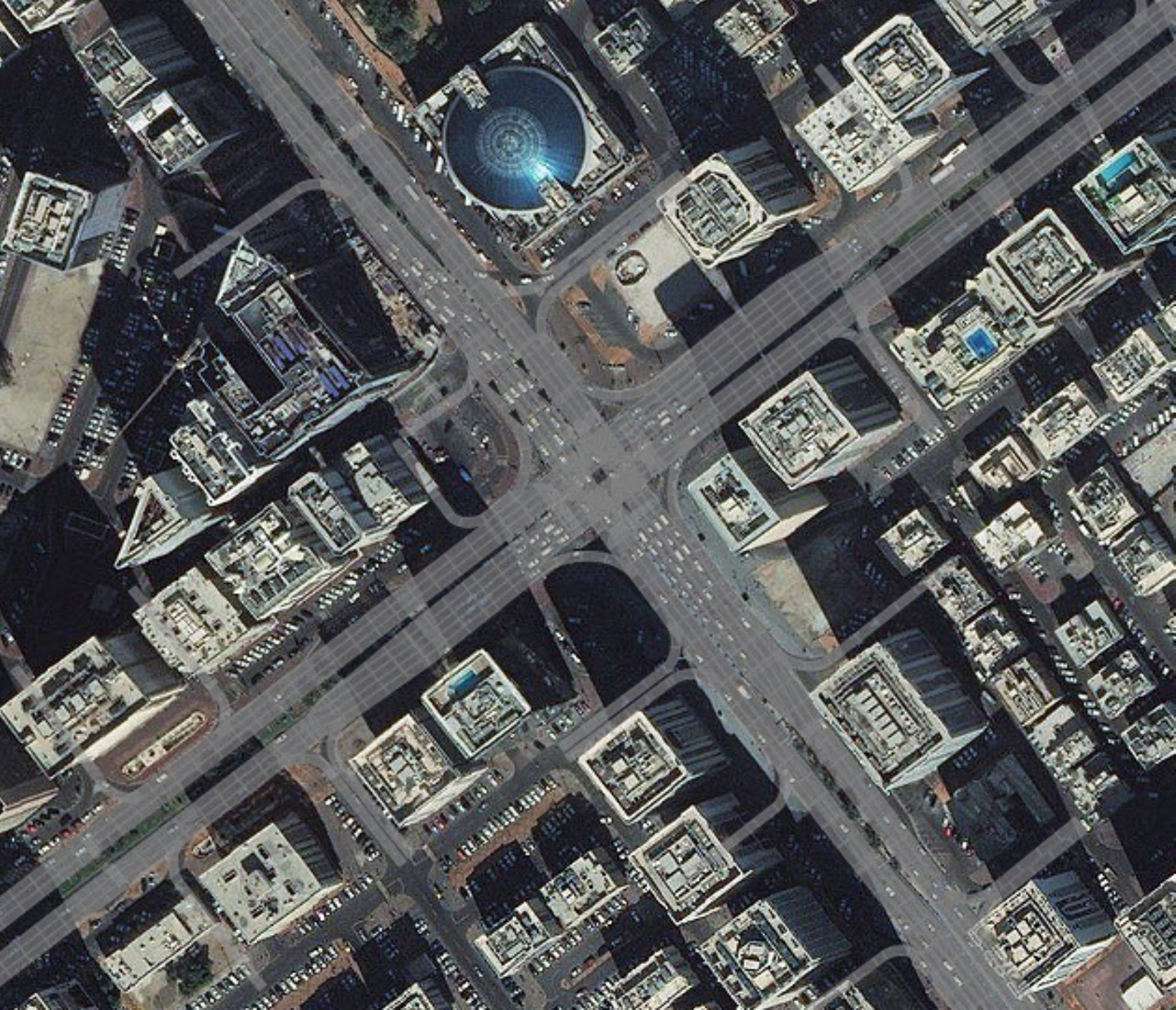}}	
	\caption{Single intersection experiment.}
	\label{F:si_map}%
\end{figure}

In the experiments below, we use a real-world isolated intersection in downtown Abu Dhabi in the United Arab Emirates (UAE), which is at the intersection of Hamdan Bin Mohammed Street and Fatima Bint Mubarak Street.  The intersection map is shown in Fig.~\ref{F:si_map}. 
\begin{figure}[ht!]
	\centering
	\subfigure[Average delay]
	{%
		\centering
		\includegraphics[width=.45\textwidth]{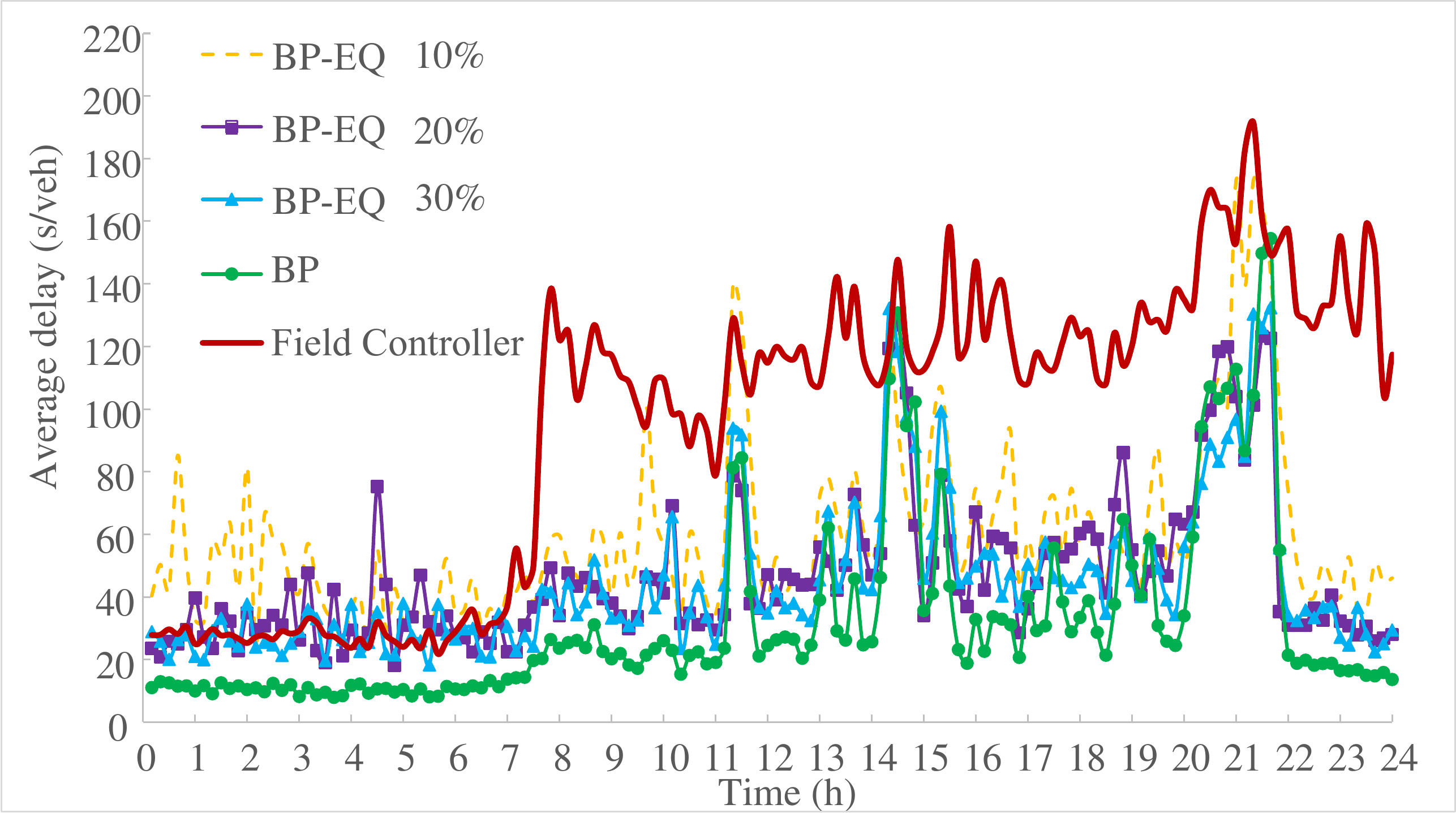}
		\label{F:si_delay}
	}%

	\subfigure[Vehicle throughput]
	{%
		\centering
		\includegraphics[width=.45\textwidth]{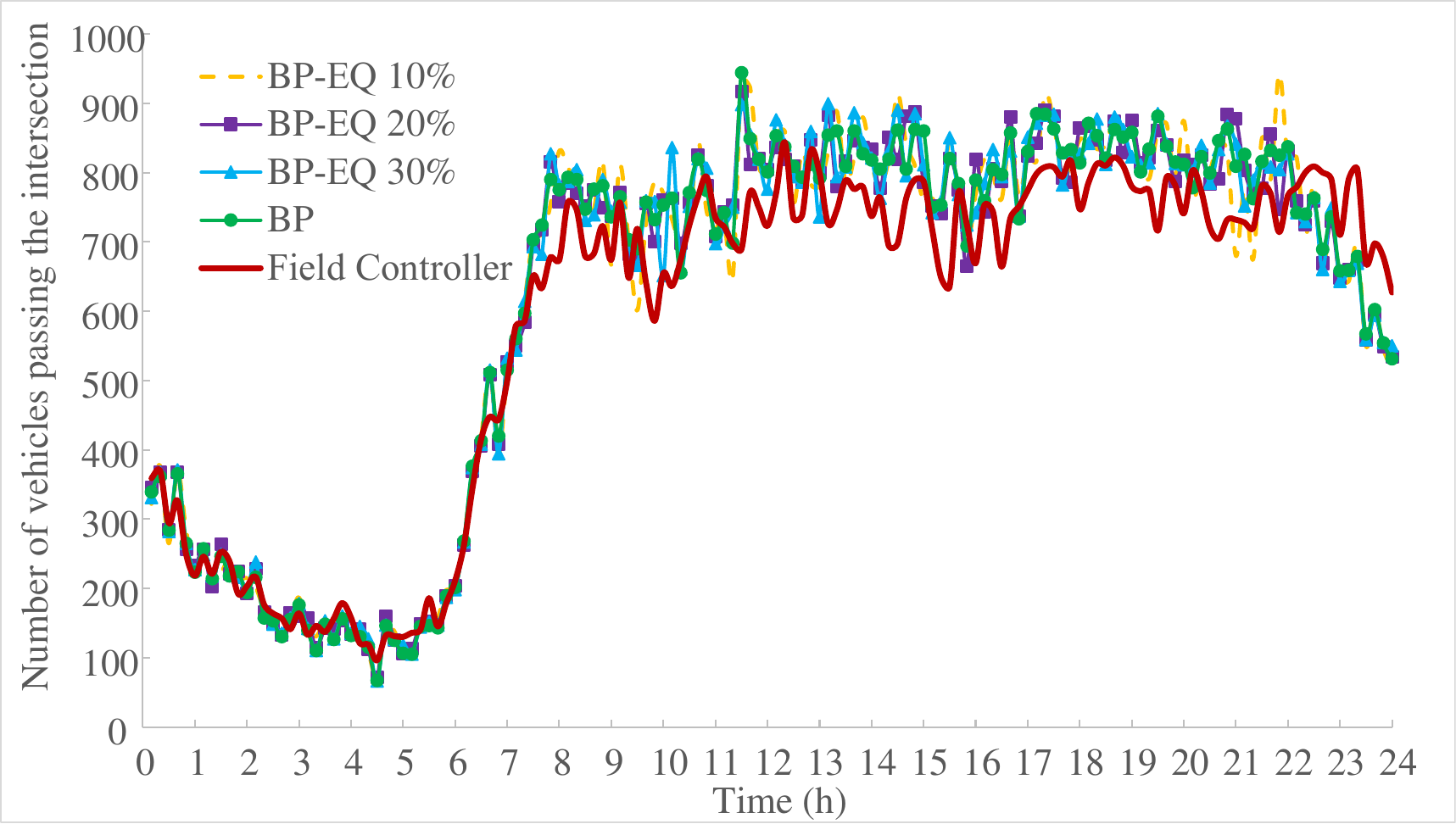}
		\label{F:si_throughput}
	}%

	\subfigure[Max-stopped-queue-length]{%
		\centering
		\includegraphics[width=.45\textwidth]{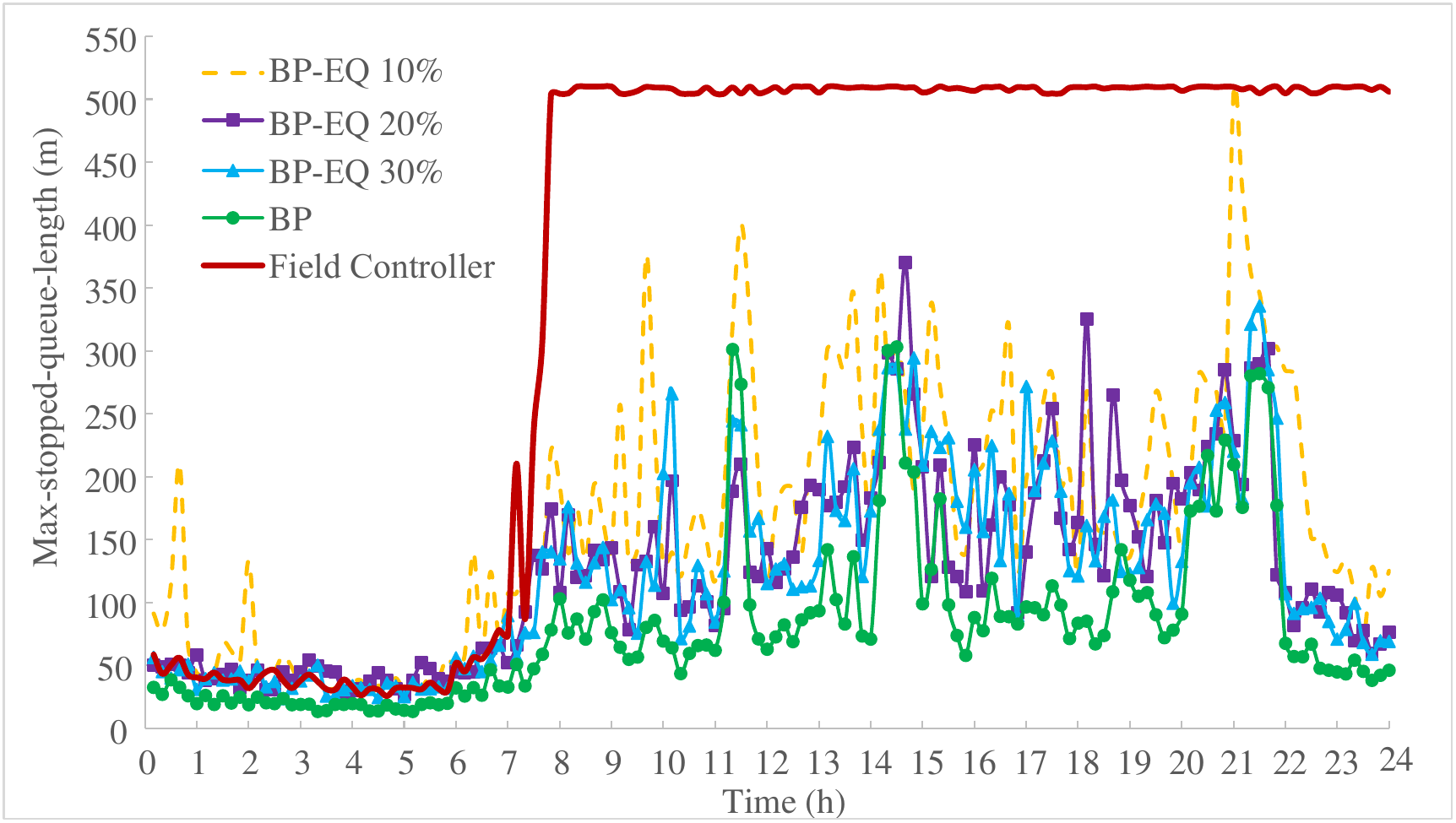}
		\label{F:si_maxqueue}
	}%
	\caption{Performance comparisons of isolated intersection between different controllers.}
	\label{F:SI_compare}
\end{figure}
We chose a typical working day, December 6, 2017, as the data collection date. High resolution traffic data were collected for the entire 24 hour period. The data were collected from loop detectors in each lane, one data point for each second of the day. The status of the traffic signal was archived by a commercial adaptive signal system that controls the intersection, also for at the second-by-second level.  The high resolution data were used to calibrate a (commercial) microscopic traffic simulation model.  Since we only calibrated an isolated intersection, we were able to use the high resolution data as input to produce a highly accurate reconstruction of the traffic dynamics at the intersection during that day (operated by an adaptive signal system). To perform our experiments, we extracted the trajectories of some of the vehicles in the (ground truth) simulation, at varying penetration levels, and treated those vehicles as CVs that share their speed and location information with the signal controller. To test the proposed control techniques, we implemented them in the simulation model via an application programming interface (API).

The comparison results are shown in Fig.~\ref{F:SI_compare}. 
We tested CVs' penetration rates of 10\%, 20\% and 30\%.  The BP-based policies were updated every 10 s. The evaluation data were collected every 10 min. Fig.~\ref{F:si_delay} compares the average vehicle delay, Fig.~\ref{F:si_throughput} compares the vehicle throughput, and Fig.~\ref{F:si_maxqueue} compares the max-stopped-queue-length (stopped vehicles in the jammed segment before the traffic light) at the different penetration rates, but also against the ground truth adaptive controller and the fully connected (100\% penetration) scenario, which we simply refer to as BP in the figures.  As can be seen, BP outperforms all other control scenarios throughout the day in terms of all three performance indices. Note that the huge gap between time 0 and 24 in average delay and max-stopped-queue-length is an absence of traffic information at time 0 (i.e., from the previous day). In reality, vehicles arriving at time 0 would encounter existing traffic, something that our simulation misses due to lack of data from the previous day (i.e., we initialize our simulations with empty networks).

According to the simulation results, when the demand is relatively low (during the late night hours, from 00:00 to 07:00), we see that the ground truth control beats BP-EQ with CV penetration rates of 20\% and less.  As mentioned in Sec.~\ref{ss:v2rho}, for any equilibrium speed-density relation, free-flow speeds are associated with a range of traffic densities and estimated queue lengths are prone to error as a result. This explains why BP-EQ performs worse than the ground truth controller in low density cases. In general, the performance of BP-EQ drops under low penetration rates in low demand scenarios in comparison to other techniques.

As the demand increases (during the day, from 07:00 to 24:00), the ground truth adaptive controller starts to perform poorly: it produces heavy congestion and its max-stopped-queue-length reaches 500 m and remains there for the duration of the congested period.  We note that (i) this is ground truth data, (ii) the max-stopped-queue-length shown in Fig.~\ref{F:si_maxqueue} is not to be interpreted as a time series of stopped queue length, but rather as a time series of maximum stopped queue lengths over 10-minute intervals (i.e., multiple signal cycles), and (iii) this is the maximum stopped queue length collected from all approaches, meaning that such max-stopped-queue-length appears on at least one (but not necessarily all) approach(es). This is the reason why the average delay and vehicle throughput do not collapse while the max-stopped-queue-length remains 500 m. During the day, where some congestion exists, BP-EQ outperforms the ground truth controller at all penetration rates. The reasons behind the performance of BP-EQ (vs. ground truth control) with a CV penetration rate that is as low as 10\% is that 1) the Newell-Franklin speed-density relation is more realistic in high density cases, and 2) more data becomes available in heavier traffic and the queue length estimation procedure produces better inputs for BP.  In general, the performance of BP-EQ improves as the CV penetration rates increase, but the marginal benefits keep decreasing; the biggest gain is obtained when the penetration rate increases from 10\% to 20\%. This implies that in high demand scenarios, a CV penetration rate of 20\% is sufficient for BP-EQ to perform well compared to a 100\% penetration rate.

\subsection{Traffic Network}

\begin{figure}[ht!]%
	\centering
	\resizebox{0.45\textwidth}{!}{%
		\includegraphics{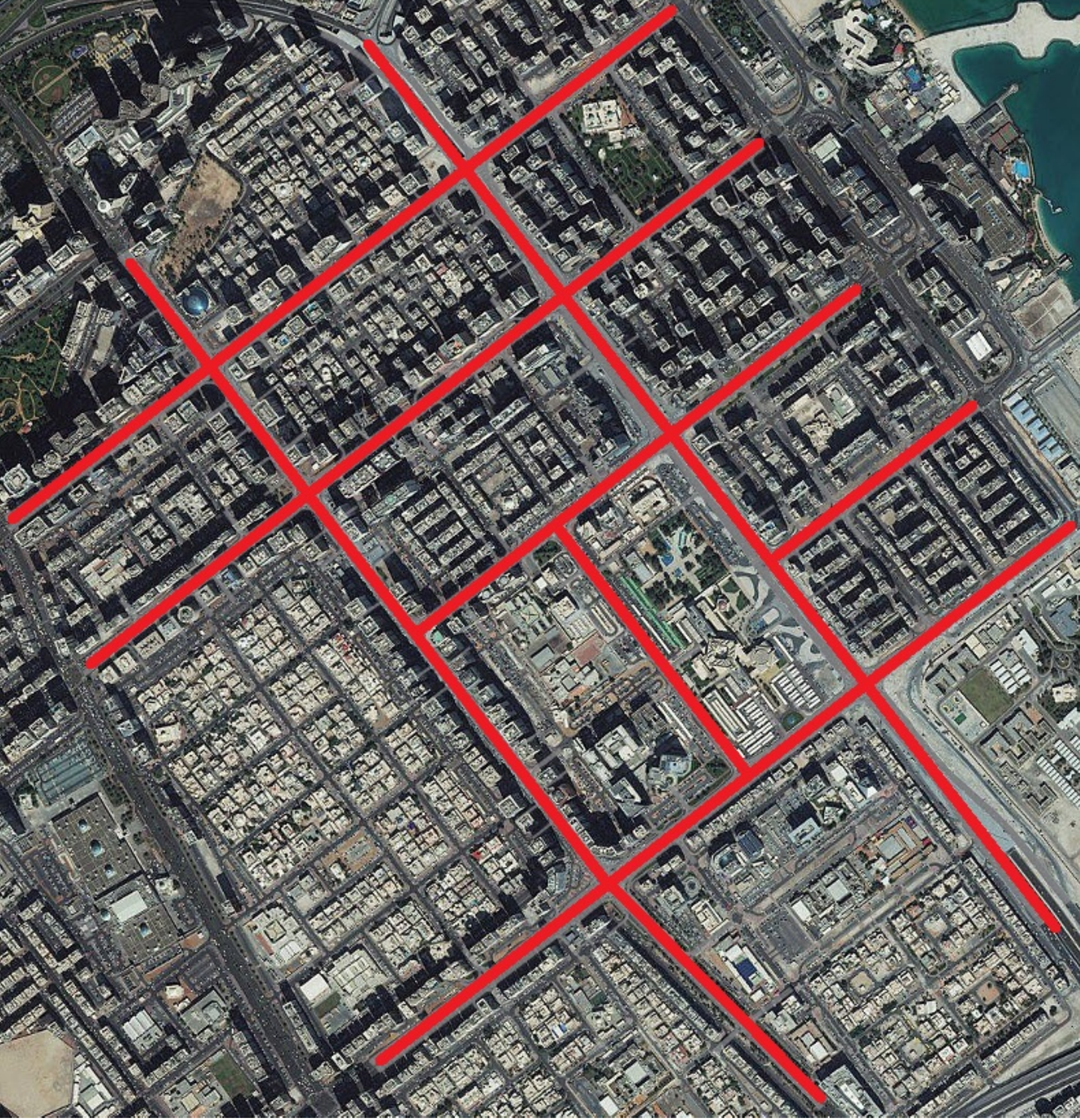}}	
	\caption{Network experiment.}
	\label{F:nw_map}%
\end{figure}

Similar to the isolated intersection, we used high resolution data to calibrate the external traffic demands of an 11-intersection network. The network map is shown in Fig.~\ref{F:nw_map}. 
Due to the absence of detectors on minor roads within the network, we have no knowledge of the trips produced or attracted to the interior links. For this reason, we do not compare the BP-based policies with the commercial adaptive signal system that controls the intersection in field, but instead optimize a fixed timing (FT) schedule (including coordination among intersections) under the calibrated external demands using simulation tools and compare it with BP.

\begin{figure}[ht!]
	\centering
	\subfigure[Average delay]
	{%
		\centering
		\includegraphics[width=.45\textwidth]{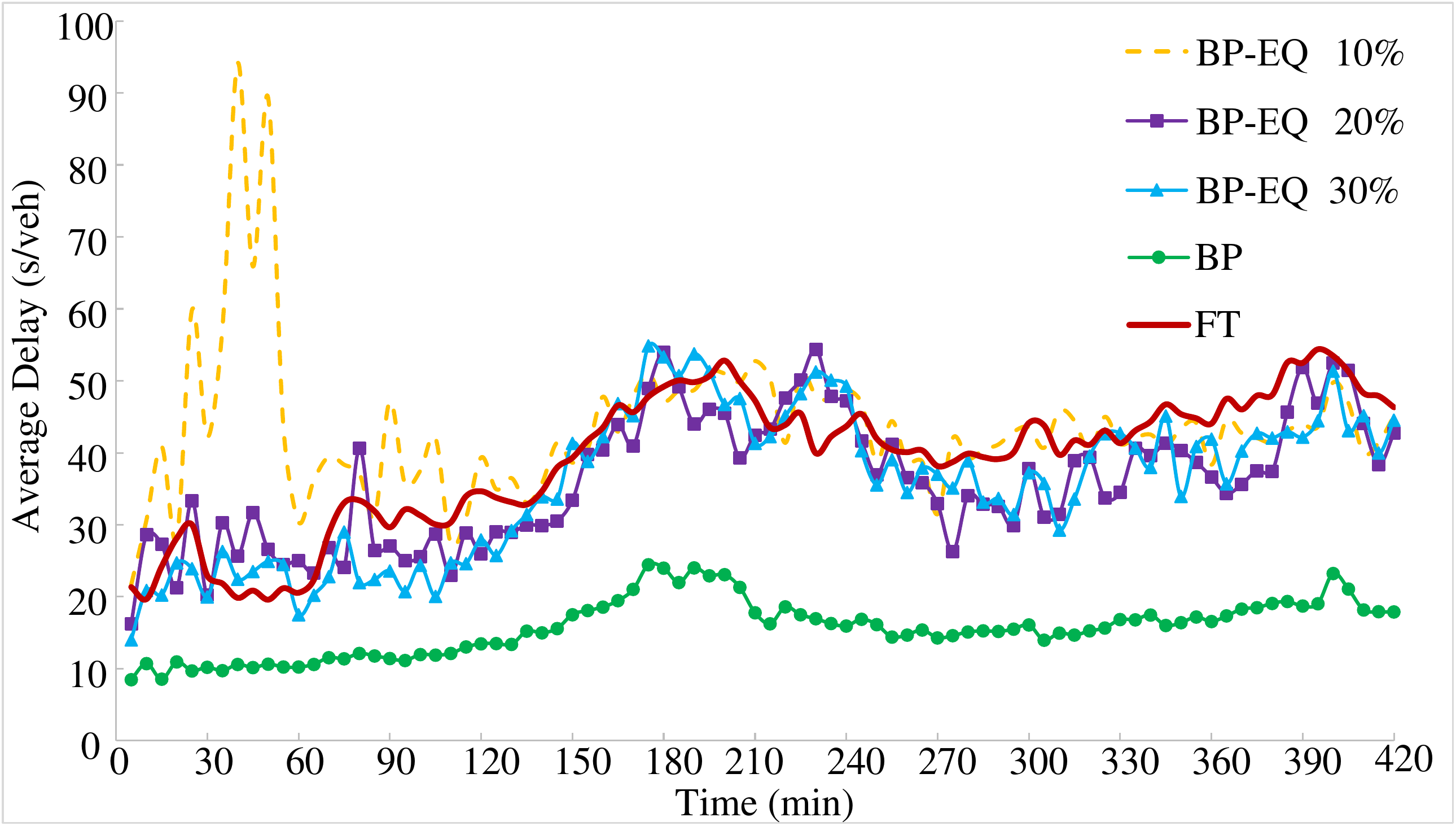}
		\label{F:nw_delay}
	}%
	
	\subfigure[Vehicle throughput]
	{%
		\centering
		\includegraphics[width=.45\textwidth]{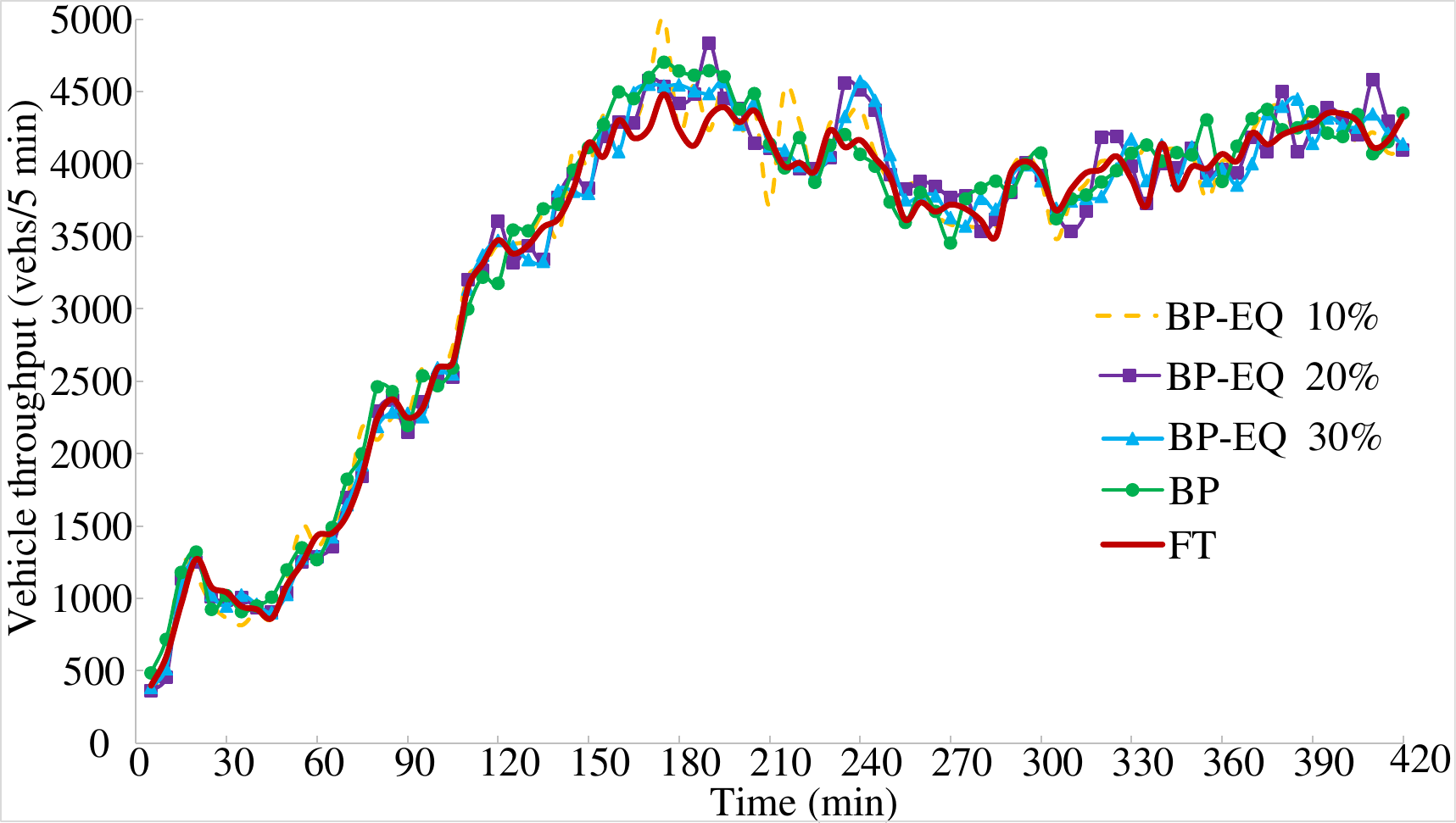}
		\label{F:nw_throughput}
	}%
	
	\subfigure[Max-stopped-queue-length]{%
		\centering
		\includegraphics[width=.45\textwidth]{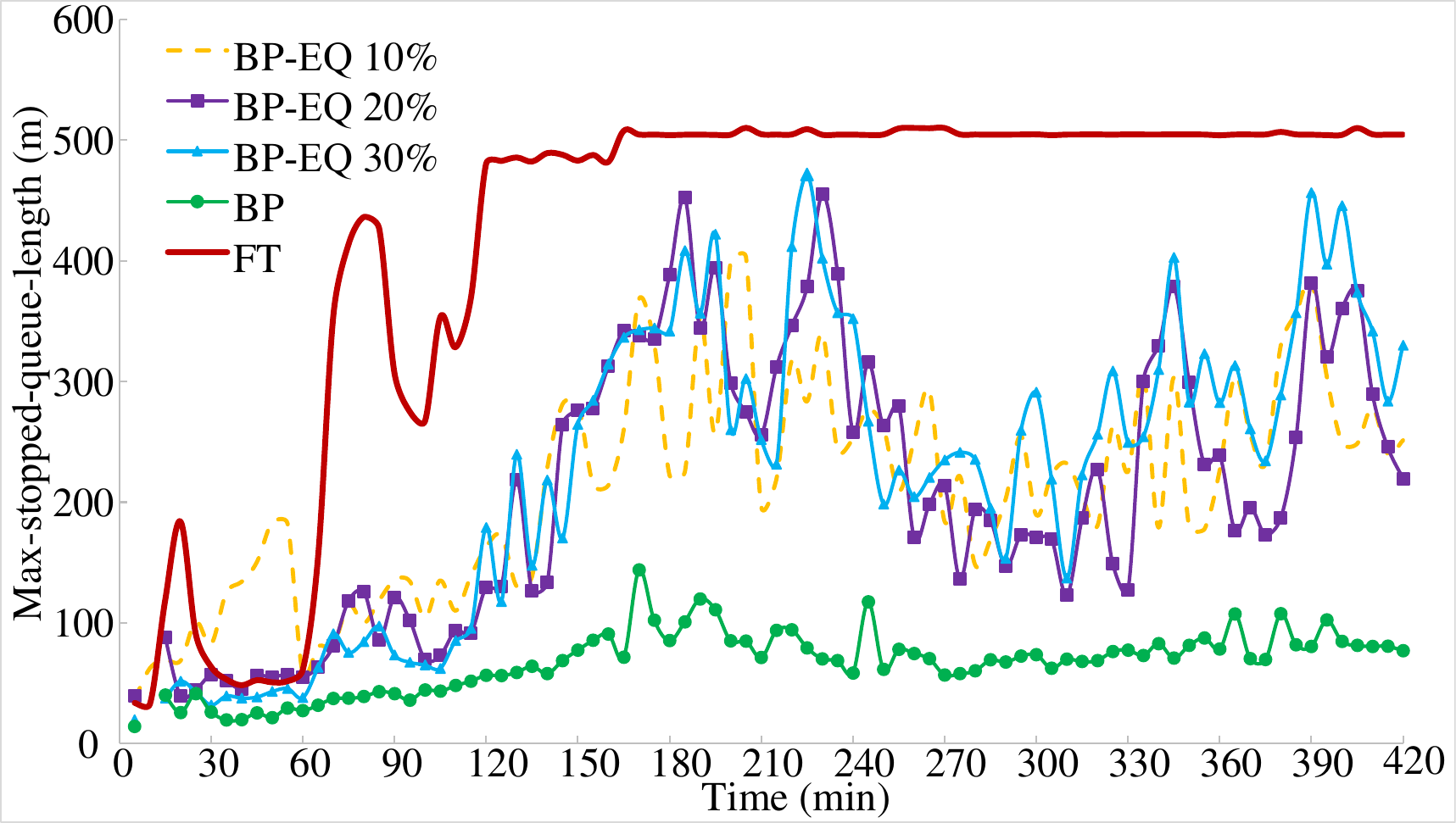}
		\label{F:nw_maxqueue}
	}%
	\caption{Performance comparisons of traffic network between different controllers.}
	\label{F:network_compare}
\end{figure}

We run 7-hour simulations with demand data calibrated from 05:00 am to 12:00 pm, on December 6, 2017.
The BP-based controls were updated every 10 s. The evaluation data were collected every 5 min. Fig.~\ref{F:nw_delay} compares the average vehicle delay, Fig.~\ref{F:nw_throughput} compares the vehicle throughput, and Fig.~\ref{F:nw_maxqueue} compares the max-stopped-queue-length at different penetration rates, but also against the FT controller and the BP with perfect knowledge.

Again, BP outperforms all other controllers in terms of delay, throughput and max-stopped-queue-length. The FT controller is able to handle the low demand cases, while the max-stopped-queue-length reaches 500 m in the high demand cases. For the BP-EQ, in the low demand scenarios, 10\% penetration of connected vehicles is too low to have a good performance. The delay increases sharply and fluctuates dramatically. However, with an increase in the demand, the number of connected vehicles increases and the 10\% CV penetration rate begins to perform better than the FT controller. With a 20\% CV penetration rate, the max-stopped-queue-length of BP-EQ remains shorter than the FT under low demand cases, and its average delay  fluctuates around the FT's average delay more gently. As the demand increases to high levels, the performance of BP-EQ becomes better than FT in terms of both average delay and max-stopped-queue-length.

\section{Conclusion and outlook}
\label{S:Conc}

Backpressure or (max weight) control has received a great deal of recent attention in the literature for many reasons including (i) the network-wide stability capabilities of BP, (ii) the fact that BP does not require knowledge of demand arrival rate to operate, and (iii) it is computationally light-weight and can scale to large networks.  BP is essentially an adaptive control technique; it has the advantage of being an open box tool that has been vetted by many researchers in recent years.

A main drawback of BP is that it requires accurate knowledge of queue lengths, which are typically difficult to collect in real-world settings (in today's road traffic systems).  Queue lengths have always been assumed to be known exactly (by assuming a fully connected environment) in the literature.  Many variants of BP control have been proposed for decentralized signal control, but none have investigated the impact of incomplete data (e.g., in a partially connected environment) on the performance of BP.  This was the question that this paper set out to investigate.

In order to investigate how BP performs with inaccurate queue lengths, we proposed a BP with estimated queue length (BP-EQ) for partially connected environments in this paper. By utilizing speed data made available by some vehicles, we estimate the speed profile for the entire lengths of the roads, and then calculate traffic densities from the estimated speeds using the Newell-Franklin speed-density relation.  The estimated densities are then used to calculate queue sizes, which serve as inputs for BP control. We tested the proposed BP-EQ control technique under different penetration rates in microscopic simulation models with calibrated data from the real world. For the isolated intersection experiments, comparisons are made with a commercial adaptive controller, used to control the intersection in the real world, as well as BP with perfect knowledge. For the traffic network experiments, comparisons are made with an optimized fixed timing controller, as well as BP with perfect knowledge. Our results show that for an isolated intersection, BP-EQ with CV penetration rates of 30\% can perform as well as the real-world adaptive controller under low demands; BP-EQ with CV penetration rates as low as 10\% outperformed the real-world adaptive controller under high demand scenarios. For the 11-intersection network, a 20\% of CV penetration rate is sufficient for BP-EQ to outperform the fixed timing controller in high demand scenarios, and have similar performance with the fixed timing controller in low demand scenarios.

Extensions to the present work as future research can follow several directions.  We utilized a simple BP for the present work and demonstrated that it remains viable with imperfect information.  One can consider improvements to the BP, utilizing more advanced variants, such as the position-weighted BP.  A more sophisticated estimation technique can also be considered for future research and combined estimation and control approaches can be explored as well.  Another aspect that can be considered for future research is online parameter selection updating and the computation complexity of the entire system. In addition, one could also seek to combine the network mobility management (e.g. \citep{li2020user}) with BP signal control to better manage the traffic.

\section*{Acknowledgment}
This work was supported by the NYUAD Center for Interacting Urban Networks (CITIES), funded by Tamkeen under the NYUAD Research Institute Award CG001 and by the Swiss Re Institute under the Quantum Cities\textsuperscript{TM} initiative.  The authors would also like to acknowledge in-kind support received from the Abu Dhabi Department of Transportation, in the form of the high-resolution traffic data that were used in our experiments.

%\section*{Resources}
%The simulation model and the trigger selection methodology will be open sourced as benchmarks for defense development.

\appendix
\gdef\thesection{Appendix \Alph{section}}

%\section*{References}
	
	%% References
	
	
\bibliographystyle{plainnat}
\bibliography{refs}
	
	
	
	
	
	
	% \end{thebibliography}

\end{document}